\documentclass[a4paper,11pt]{article}
\pdfoutput=1 

\usepackage{jheppub} 

\usepackage[T1]{fontenc} 

\title{\boldmath The spectrum of Hawking radiation in Tsallis statistical mechanics}

\author{Yang Liu}

\affiliation[a]{School of Physics and Astronomy, University of Nottingham, Nottingham NG7 2RD, UK}
\affiliation[b]{Nottingham Centre of Gravity, University of Nottingham, Nottingham NG7 2RD, UK}

\emailAdd{yang.liu@nottingham.ac.uk}

\abstract{Hawking radiation is one of the cores in modern gratitational theory. Several articles have calculated the spectrum of Hawking radiation in Boltzmann-Gibbs statistical mechanics. However, based on recent researches, gravitational systems cannot be studied by the standard statistical mechanics. In this article, we calculate the modification to the spectrum of Hawking radiation in Tsallis statistical mechanics. We obtain the modified Stefan-Boltzmann's law and modified power of Hawking radiation. We confirm the conclusion proposed by Giddings, namely, the radiation should originate from the effective radius, which extends well outside the horizon of black-hole. The lifetime of black hole and the effect of large $q$ are discussed as well. }

\begin{document} 
\maketitle
\flushbottom

\section{Introduction}
\label{sec:intro}
In Hawking's breakthrough work in 1974 [1], he showed that black holes can emit particles spontaneously at a temperature which is inversely proportional to their mass. Black holes have been becoming extremely important in classical and quantum gravity theories since then [2]. Meanwhile, a series of works devoted to calculating the Hawking radiation spectrum and temperature to explore a possible theory of quantum gravity [3-5]. One of the most important problems about Hawking radiation is that where the radiation originates? A common picture is that it arises from excitations very near or at the horizon $r_H$ [6], while Giddings [6,7] proposed that the Hawking black-hole radiation spectrum originates from an effective quantum “atmosphere” $r_A$ which extends well outside the horizon $r_H$. \\
For a $(D+1)$-dimensional black hole, the semi-classical Hawking radiation power for one bosonic degree of freedom in standard Gibbs-Boltzmann statistical mechanics is given by [7,8,9,10]
\begin{equation}\label{eq:1.1}
P_{BH} = \frac{1}{2^{D-1} \pi^{D/2} \Gamma(D/2)} \sum_j \int_{0}^{\infty} \Gamma \frac{\omega^D}{\exp(\omega/T_{BH}) -1} d\omega, 
\end{equation}
where $\omega$ is the emitted frequency of field, $j$ is the angular harmonic index of the emitted field modes and $\Gamma=\Gamma(\omega;j,D)$ is the greybody factor. However, as some refs.[11,12,13,14] pointed out that Boltzmann-Gibbs statistical mechanics cannot be applied to study gravitational systems. Therefore, the spectrum $(1.1)$ should be modified. Tsallis generalized standard statistical mechanics (which arises from the hypothesis of weak probabilistic correlations and their connection to ergodicity) to nonextensive one, which can be applied in all cases, and still possessing standard Boltzmann-Gibbs theory as a limit [12,14]. A natural question is that how nonextensive statistical mechanics modifies the spectrum of Hawking radiation? In our previous paper [14], we have considered the effect of Rényi entropy and proposed that Hawking radiation should originate from the effective radius $r_A$ instead of horizon $r_H$, which is the same as Giddings' proposal in finite dimension $D$. However, in that article, we still used Boltzmann-Gibbs statiscal mechanics [14]. In this article, we will apply Tsallis statistical mechanics to study the modified Hawking radiation spectrum. \\
Here we briefly review the basics of Tsallis statistical mechanics. The generalized form of entropy in Tsallis statistical mechanics is [12] 
\begin{equation}\label{eq:1.2}
S_q = k_B \frac{1- \sum_{i=1}^{W} p^q_i}{q-1}, \qquad q \in \textbf{R},
\end{equation}
where $k_B$ is the Boltzmann constant, $W$ is the total number of physical states of the system, $q$ is a real parameter called the non-extensivity parameter and the set of probabilities $p_i$ satisfies
\begin{equation}\label{eq:1.3}
\sum_{i=1}^{W} p_i =1.
\end{equation}
Eq.$(1.2)$ can be recovered to Boltzmann-Gibbs form if we take the limit $q \rightarrow 1$, namely, [12],
\begin{equation}\label{eq:1.4}
\lim_{q \rightarrow 1} S_q = -k_B \sum_{i=1}^{W} p_i \ln p_i.
\end{equation}
According to ref.[15], the most probable distribution over a single state in nonextensive statistical mechanics is given by:
\begin{equation}\label{eq:1.5}
\bar{n}(g,q) = \frac{1}{[1-(1-q)\beta (\epsilon_k - \mu)]^{1/(q-1)} +2g-1},
\end{equation}
where $g$ is a statistic number, $\beta = 1/(k_B T)$, and $\mu$ is the chemical potential. Obviously, when $q \rightarrow 1$, Bose-Einstein distribution for $g=0$ and Fermi-Dirac distribution for $g=1$ can be recovered [15].\\
Since the photon number in photon is not conservative, we have $\mu=0$. Photon is boson, thence $g=0$. Considering the energy of photon is given by $\epsilon_k = \hbar \omega_k$, where $\omega_k$ is the frequency of photon, then the most probable distribution of photon gas should be rewriten as:
\begin{equation}\label{eq:1.6}
\bar{n}(0,q) = \frac{1}{[1-(1-q)\beta \hbar \omega ]^{1/(q-1)} -1}.
\end{equation} \\
This article is composed as follows: in section 2, we obtain the modified Stefan-Boltzmann's law in $(D+1)$-dimension. In section 3, we calculate the modified Hawking radiation power in $(D+1)$-dimension. In section 4, the effective radius and lifetime of black hole are discussed. In section 5, we briefly comment on the effect of large $q$. The conclusions and outlook are discussed in section 6. We take $G=c=k_B=\hbar=1$.   

\section{Modified Stefan-Boltzmann's law in (D+1)-dimension}
In ref.[6], Giddings has pointed out that we can define the effective radius $r_A$ of the black-hole quantum atmosphere by equating the Hawking radiation power $P_{BH}$ of the emitting black hole with the corresponding $P_{BB}$ of a flat space perfect balckbody emitter [6,7]. The scalar radiation power of a spherically-symmetric blackbody (BB) of temperature $T$ and radius $R$ in $(D+1)$-dimension spacetime is given by the Stefan-Boltzmann law:
\begin{equation}\label{eq:2.1}
P_{BB}=\sigma A_{D-1} (R) T^{D+1},
\end{equation}
where
\begin{equation}\label{eq:2.2}
\sigma = \frac{D\Gamma(D/2) \zeta(D+1)}{2 \pi^{D/2+1}}
\end{equation}
is the Stefan-Boltzmann constant in $(D+1)$-dimension and
\begin{equation}\label{eq:2.3}
A_{D-1}(R)= \frac{2\pi^{D/2}}{\Gamma(D/2)} R^{D-1}
\end{equation} 
is the surface area of the emitting body in $(D+1)$-dimension [7].\\ 
In this section we will derive the modified Stefan-Boltzmann law and obtain the modified Stefan-Boltzmann constant $\sigma_q$ in $(D+1)$-dimension. Based on the derivation of Stefan-Boltzmann's law [15], the integral we need to consider is
\begin{equation}\label{eq:2.4}
I_q(D)= \int_{0}^{\infty} dx \frac{x^D}{[1-(1-q)x]^{1/(q-1)}-1},
\end{equation}
where $x=\beta \omega$. In fact, the only difference is that the denominator $\exp(x)-1$ in the integral should be replaced by $[1-(1-q)x]^{1/(q-1)}-1$. We can expect that the modified blackbody radiation power should be equal to $\sigma_q A_{D-1} (R) T^{D+1}$, where $\sigma_q$ is the modified Stefan-Boltzmann constant since nonextensive statistical mechanics has no effect on $A$ and $T$. Now we will consider the integral for the case of $q<1$ and $q>1$, respectively. Here we follow the method given in ref.[15]. \\
The definition of the gamma function is 
\begin{equation}\label{eq:2.5}
\Gamma (\alpha) = \int_{0}^{\infty} t^{\alpha -1} e^{-t} dt, \quad \alpha > 0. 
\end{equation}
If we substitute $t= [1- (1-q)x]v$, then we have
\begin{equation}\label{eq:2.6}
[1-(1-q)x]^{1/(q-1)} = \{1/\Gamma(\frac{1}{1-q}) \} \int_{0}^{\infty} v^{[1/(q-1)]-1} e^{-v} e^{v(1-q)x}dv,
\end{equation}
where $\alpha=1/(1-q)$ and $\alpha >0$ since $q<1$.\\ 
Based on the definition of the gamma function $\Gamma$, we have the following identity:
\begin{equation}\label{eq:2.7}
1 \equiv \{1/\Gamma(\frac{1}{1-q}) \} \int_{0}^{\infty} v^{[1/(1-q)]-1} e^{-v} dv.
\end{equation}
Combine eq.$(2.6)$ and $(2.7)$, we can obtain
\begin{equation}\label{eq:2.8}
[1-(1-q)x]^{1/(q-1)} -1 = \{1/\Gamma(\frac{1}{1-q}) \} \int_{0}^{\infty} v^{[1/(1-q)]-1} e^{-v} (e^{v(1-q)x}-1) dv.
\end{equation}
Substituting eq.$(2.8)$ into eq.$(2.4)$, then we have
\begin{equation}\label{eq:2.9}
I_q(D) = \{1/\Gamma(\frac{1}{1-q}) \int_{0}^{\infty} v^{[1/(1-q)]-1} e^{-v} \} \int_{0}^{\infty} dx \frac{x^D}{e^{v(1-q)x} -1}.
\end{equation}
Furthermore, we define $y \equiv v(1-q)x$, then eq.$(2.9)$ can be rewritten as
\begin{equation}\label{eq:2.10}
I_q(D) = \{\Gamma (\frac{1}{1-q}) / (1-q)^{D+1} \times \int_{0}^{\infty} dv v^{[1/(q-1)]+D} e^{-v} \} \int_{0}^{\infty} dy \frac{y^D}{e^y -1}.
\end{equation}
From eq.$(2.7)$, we can rewrite
\begin{equation}\label{eq:2.11}
\int_{0}^{\infty} dv v^{[1/(q-1)]+D} e^{-v} \Gamma(\frac{1}{1-q} +D+1),
\end{equation}
then for $q<1$, 
\begin{equation}\label{eq:2.12}
I_q(D) = \{\Gamma (\frac{1}{1-q}) / (1-q)^{D+1} \times \Gamma(\frac{1}{1-q} +D+1) \} \Gamma(D+1) \zeta (D+1),
\end{equation}
where we have used 
\begin{equation}\label{eq:2.13}
\int_{0}^{\infty} dy \frac{y^D}{e^y -1} = \Gamma(D+1) \zeta (D+1).
\end{equation}
Since $\Gamma(z+1) = z\Gamma(z)$, then eq.$(2.13)$ can be written as
\begin{equation}\label{eq:2.14}
I_q(D) = \{\frac{1}{(1-q)^{D+1} (\frac{1}{1-q} + D) (\frac{1}{1-q} + D-1) ... (\frac{1}{1-q})} \} \Gamma(D+1) \zeta (D+1), \quad q<1.
\end{equation}
Now we will calculate the integral $(2.4)$ for $q>1$. Based on ref.[15], we know that 
\begin{equation}\label{eq:2.15}
\Gamma(\alpha) \frac{i}{2\pi} \oint_C dv (-v)^{-q/(q-1)} e^{-v} =1,
\end{equation}
where $\alpha >0$. If we substitute
\begin{equation}\label{eq:2.16}
t=v[1-(1-q)x],
\end{equation}
we have
\begin{equation}\label{eq:2.17}
[1-(1-q)x]^{1/(q-1)} = \frac{i}{2\pi} \Gamma(\frac{q}{q-1}) \oint_C dv (-v)^{-q/(q-1)} e^{-v} e^{-v(q-1)x},
\end{equation}
where $\alpha = q/(q-1)>0$ since $q>1$. On the other hand,
\begin{equation}\label{eq:2.18}
1= \frac{i}{2\pi} \Gamma(\frac{q}{q-1}) \oint_C dv (-v)^{-q/(q-1)} e^{-v}. 
\end{equation}
Combine eq.$(2.16)$ and $(2.17)$, one can obtain
\begin{equation}\label{eq:2.19}
[1-(1-q)x]^{1/(q-1)} -1 = \frac{i}{2\pi} \Gamma(\frac{q}{q-1}) \oint_C dv (-v)^{-q/(q-1)} e^{-v} (e^{-v(q-1)x} -1).
\end{equation}
Therefore, the integral $(2.4)$ should be
\begin{equation}\label{eq:2.20}
I_q(D)= \{1/\Gamma(\frac{q}{q-1}) \times \frac{i}{2\pi} \oint_C dv (-v)^{-q/(q-1)} e^{-v} \} \int_{0}^{\infty} \frac{x^D}{e^{-v(q-1)x}-1}.
\end{equation}
If we change the variable of integration 
\begin{equation}\label{eq:2.21}
y=v(1-q)x,
\end{equation}
and consider eq.$(2.15)$, then the integral $(2.4)$ is
\begin{equation}\label{eq:2.22}
I_q(D) = \{\frac{1}{(q-1)^{D+1} (\frac{q}{q-1} -1) (\frac{q}{q-1}-2) ... (\frac{q}{q-1} -D-1)} \} \Gamma(D+1) \zeta (D+1), \quad q>1.
\end{equation}
Furthermore, the integral $(2.4)$ can be simplified
\begin{equation}\label{eq:2.23}
I_q(D) = \{\frac{1}{(D+1-Dq) (D-(D-1)q)... (2-q)} \} \Gamma(D+1) \zeta (D+1),  
\end{equation}
for $q<1$ and $q>1$. Hence, the modified Stefan-Boltzmann constant $\sigma_q$ is given by
\begin{equation}\label{eq:2.24}
\sigma_q(D) = \{\frac{1}{(D+1-Dq) (D-(D-1)q)... (2-q)} \} \sigma,  
\end{equation}
and the modified Stefan-Boltzmann law is 
\begin{equation}\label{eq:2.25}
\tilde{P}_{BB} = \sigma_q A_{D-1} (R) T^{D+1}, 
\end{equation}
where $\tilde{P}_{BB}$ is the modified blackbody radiation power. In particular, at order $O(q-1)$, we have
\begin{equation}\label{eq:2.26}
\sigma_q = (1+B(q-1))\sigma,  
\end{equation}
where $\sigma$ is the Stefan-Boltzmann constant obtained in the standard statistical mechanics and 
\begin{equation}\label{eq:2.27}
B= \frac{d}{dq} I_q(D)  
\end{equation}
around $q=1$. We list the value of $B$ for different dimension $D$ in Table 1. \\
\begin{table}
	\caption{The value of $B$ for different dimension $D$, where $B$ is the derivative of $I_q(D)$ with respect to $q$.}
	\begin{tabular}{|c c c c c c c c c|} 
		\hline
		$D+1$ & 4 & 5 & 6 & 7 & 8 & 9 & 10 & 11 \\
		\hline
		$B$ & 6 & 10 & 15 & 21 & 28 & 36 & 45 & 55 \\
		\hline
	\end{tabular}    
\end{table}

\section{Modified the power of Hawking radiation in (D+1)-dimension}
For a $(D+1)$-dimensional black hole, the semi-classical Hawking radiation power for one bosonic degree of freedom in standard statistical mechanics is given by [7,8,9,10]
\begin{equation}\label{eq:3.1}
P_{BH} = \frac{1}{2^{D-1} \pi^{D/2} \Gamma(D/2)} \sum_j \int_{0}^{\infty} \Gamma \frac{\omega^D}{\exp(\omega/T_{BH}) -1} d\omega, 
\end{equation}
where $\omega$ is the emitted frequency of field, $j$ is the angular harmonic index of the emitted field modes and $\Gamma=\Gamma(\omega;j,D)$ is the greybody factor. The temperature of the black hole is 
\begin{equation}\label{eq:3.2}
T_{BH} = \frac{(D-2)}{4\pi r_{BH}}.
\end{equation}
where $r_{BH}$ is the horizon radius of the black hole [7]. We have set $G=c=\hbar=k_B=1$. Now considering the effect of nonextensive statistical mechanics, i.e., eq.$(1.6)$, the Hawking radiation power, eq.$(3.1)$, should be modified to 
\begin{equation}\label{eq:3.3}
P_{BH} = \frac{1}{2^{D-1} \pi^{D/2} \Gamma(D/2)} \sum_j \int_{0}^{\infty} \Gamma \frac{\omega^D}{[1+(q-1)\beta \omega]^{1/(q-1)} -1} d\omega, 
\end{equation}
where $\beta = 1/T_{BH}$. In this section, we only consider the case of very small $(q-1)$. We can write
\begin{equation}\label{eq:3.4}
[1+(q-1)\beta \omega]^{1/(q-1)} = \exp(\omega/T_{BH}) + K,
\end{equation}
where $K$ is a function of order $O(q-1)$ and we will determine the form of $K$ later. If we define $f(\omega,q)= [1+(q-1)\beta \omega]^{1/(q-1)}$, then we have
\begin{equation}\label{eq:3.5}
\frac{1}{f(\omega,q) -1} = \frac{1}{\exp(\omega/T_{BH}) + K -1} \simeq \frac{1}{\exp(\omega/T_{BH}) -1} \{1-\frac{K}{\exp(\omega/T_{BH}) -1}\}.
\end{equation}
Therefore, the Hawking radiation power is
\begin{equation}\label{eq:3.6}
\begin{aligned}
P_{TS} & = \frac{1}{2^{D-1} \pi^{D/2} \Gamma(D/2)} \sum_j \int_{0}^{\infty} \Gamma \frac{\omega^D}{f(\omega,q) -1} d\omega \\
&\simeq \frac{1}{2^{D-1} \pi^{D/2} \Gamma(D/2)} \sum_j \int_{0}^{\infty} \Gamma \frac{\omega^D}{\exp(\omega/T_{BH}) -1} \{1-\frac{K}{\exp(\omega/T_{BH}) -1}\} d\omega \\
&= P_{BH} - \frac{1}{2^{D-1} \pi^{D/2} \Gamma(D/2)} \sum_j \int_{0}^{\infty} \Gamma \frac{\omega^D K}{(\exp(\omega/T_{BH}) -1)^2} d\omega,
\end{aligned}
\end{equation}
where $P_{TS}$ is the Hawking radiation considering the effect of Tsallis statistical mechanics while $P_{BH}$ is the Hawking radiation power given by the Boltzmann-Gibbs statistical mechanics, namely, eq.$(3.1)$. \\
Since $f(\omega,q)= [1+(q-1)\beta \omega]^{1/(q-1)}$, we have
\begin{equation}\label{eq:3.7}
\ln f(\omega,q) \simeq \beta \omega - \frac{1}{2} (q-1) \beta^2 \omega^2.
\end{equation}
Hence,
\begin{equation}\label{eq:3.8}
f(\omega,q) \simeq e^{\beta \omega} - \frac{1}{2} (q-1) \beta^2 \omega^2 e^{\beta \omega}.
\end{equation}
Thus we know that
\begin{equation}\label{eq:3.9}
K = - \frac{1}{2} (q-1) \beta^2 \omega^2 e^{\beta \omega},
\end{equation}
and
\begin{equation}\label{eq:3.10}
P_{TS}=P_{BH}+\Delta P= P_{BH} + \frac{(q-1)\beta^2}{2^{D} \pi^{D/2} \Gamma(D/2)} \sum_j \int_{0}^{\infty} \Gamma \frac{\omega^D  e^{\beta \omega} \omega^2}{(\exp(\beta \omega) -1)^2} d\omega,
\end{equation}
where
\begin{equation}\label{eq:3.11}
\Delta P= \frac{(q-1)\beta^2}{2^{D} \pi^{D/2} \Gamma(D/2)} \sum_j \int_{0}^{\infty} \Gamma \frac{\omega^D  e^{\beta \omega} \omega^2}{(\exp(\beta \omega) -1)^2} d\omega.
\end{equation}
When $q=1$, $P_{TS}=P_{BH}$ which recovers the standard result of Hawking radiation power. Now we rewrite $\Delta P$ in terms of $P_{BH}$ based on modified Stefan-Boltzmann's law in $(D+1)$-dimension. \\
In ref.[6,7], Giddings has pointed out that we can define the effective radius $r_A$ of the black-hole quantum atmosphere by equating the Hawking radiation power $P_{BH}$ of the emitting black hole with the corresponding $P_{BB}$ of a flat space perfect balckbody emitter [6,7]. At present, we should have 
\begin{equation}\label{eq:3.12}
P_{TS}(r_H, T_{BH})= \tilde{P}_{BB}(r_A, T_{BH}).
\end{equation}
At order $O(q-1)$, eq.$(3.12)$ becomes
\begin{equation}\label{eq:3.13}
P_{BH}(r_H, T_{BH}) + \Delta P = (1+B(q-1))P_{BB}(r_A, T_{BH}),
\end{equation}
then we can obtain the strong constraint for $\Gamma(\omega;j,D)$:
\begin{equation}\label{eq:3.14}
\sum_j \int_{0}^{\infty} \Gamma(\omega;j,D) \frac{\omega^{D+2}e^{\beta \omega}}{(e^{\beta \omega}-1)^2} d\omega = \frac{2B}{\beta^2} \sum_j \int_{0}^{\infty} \Gamma(\omega;j,D) \frac{\omega^{D}}{e^{\beta \omega}-1} d\omega.
\end{equation}
In addition, considering eq.$(3.11)$, at order $O(q-1)$, $P_{TS}$ is approximately equal to 
\begin{equation}\label{eq:3.15}
P_{TS} (r_H, T_{BH}) = (1+B(q-1)) P_{BH} (r_H, T_{BH}).
\end{equation}
In principle, we can use eq.$(3.15)$ to determine the effect of Tsallis statistics and the effect of higher orders is difficult to detect. 

\section{Effective radius and lifetime of black holes}
As we have pointed out, one can define the effective radius $r_A$ of the black-hole quantum atmosphere by equating the Hawking radiation power $P_{BH}$ of the emitting black hole with the corresponding $P_{BB}$ of a flat space perfect balckbody emitter [6,7]. Considering the effect of Tsallis statistical mechanics, we have  
\begin{equation}\label{eq:4.1}
P_{TS}(r_H, T_{BH})= \tilde{P}_{BB}(\tilde{r}_A, T_{BH}),
\end{equation}
where $P_{TS}$ and $\tilde{P}_{BB}$ are the modified Hawking radiation power of a black hole and the modified blackbody radiation power of a flat space perfect balckbody emitter considering Tsallis statistical mechanics. Combining eq.$(2.1)$, $(2.2)$, $(2.3)$, $(3.2)$ and $(4.1)$, we can obtain
\begin{equation}\label{eq:4.2}
\tilde{r}_A = [(D+1-Dq)(D-(D-1)q)...(2-q)]^{\frac{1}{D-1}} [\frac{\pi}{D \zeta(D+1)} (\frac{4\pi}{D-2})^{D+1} \bar{P}_{TS}]^{\frac{1}{D-1}} \times r_H,
\end{equation}
where $\tilde{r}_A$ is the effective radius in Tsallis statistical mechanics and 
\begin{equation}\label{eq:4.3}
\bar{P}_{TS} = P_{TS} \times r^2_H.
\end{equation}
At order $O(q-1)$, we can find that
\begin{equation}\label{eq:4.4}
\tilde{r}_A =r_A,
\end{equation}
where 
\begin{equation}\label{eq:4.5}
r_A = [\frac{\pi}{D \zeta(D+1)} (\frac{4\pi}{D-2})^{D+1} \bar{P}_{BH}]^{\frac{1}{D-1}} \times r_H
\end{equation}
is the effective radius obtained in standard statistical mechanics [7]. Based on ref.[7], we know that the Hawking black-hole radiation spectrum originates from an effective quantum “atmosphere” which extends well outside the black-hole horizon, namely, $r_A$ is far from horizon of black-hole. From eq.$(4.4)$, we know that the conclusion keeps at order $O(q-1)$. \\  
Moreover, we calculate the lifetime of black hole. According to energy conservation, we have
\begin{equation}\label{eq:4.6}
\frac{d}{dt} M = -P_{TS} = - \tilde{P}_{BB}.
\end{equation}
In $(3+1)$-dimension, we know that $A=4\pi r^2_A$ and $r_A = 2.679 r_H$ [7]. In addition, we have $M = \frac{1}{2} r_H$. Therefore, from eq.$(4.6)$, we obtain that
\begin{equation}\label{eq:4.7}
t = 1.488 \times \frac{\pi^3 r^3_H}{\sigma_q}.
\end{equation}
From eq.$(2.24)$, then eq.$(4.7)$ can be rewritten as
\begin{equation}\label{eq:4.8}
t = (4-3q)(3-2q)(2-q)t_0,
\end{equation}
where 
\begin{equation}\label{eq:4.9}
t_0 = 1.488 \times \frac{\pi^3 r^3_H}{\sigma}.
\end{equation}
According to ref.[16], we know that $0.88<q<1.05$ from observations. Hence, the lifetime of black hole is
\begin{equation}\label{eq:4.10}
0.727t_0 < t < 1.889 t_0.
\end{equation}
In principle, we can compare $(4.10)$ with $t_0$ to detect the effect of Tsallis statistical mechanics in the future. \\

\section{Effect of large q}
In this section, we briefly consider the effect of very large $q$. When $q \rightarrow \infty$, we know 
\begin{equation}\label{eq:5.1}
\lim_{q \rightarrow \infty} \ln [1+(q-1)\beta_H \omega]^{1/(q-1)} = \frac{1}{q}.
\end{equation}
Therefore,
\begin{equation}\label{eq:5.2}
\lim_{q \rightarrow \infty} [1+(q-1)\beta_H \omega]^{1/(q-1)} = \lim_{q \rightarrow \infty} \frac{1}{e^{1/q}-1} = q.
\end{equation} 
Furthermore, from $(2.4)$, we know that when $q \rightarrow \infty$
\begin{equation}\label{eq:5.3}
\sigma_q = \frac{\sigma}{(-1)^D D! q^D}.
\end{equation}
Hence, we know that $D$ must be even. And the effective radius $r_A$ should be
\begin{equation}\label{eq:5.4}
r_A = [\frac{(-1)^D D! q^D \pi}{D\zeta(D+1)} (\frac{4\pi}{D-2})^{D+1} \bar{P}_{TS}]^{1/(D-1)} \times r_H,
\end{equation}
where $\bar{P}_{TS}$ has been given by $(4.3)$.

\section{Conclusions and outlook}
In his breakthrough work in 1974 [1], Stefen Hawking showed that black holes can emit particles spontaneously. Black holes have been becoming extremely important in classical and quantum gravity theories since then [2]. Researchers have studied the spectrum of Hawking radiation [3,4,5,8,9,10]. In these articles people only considered Boltzmann-Gibbs statistical mechanics which cannot be applied to study gravitational systems. Therefore, one direct question is to consider the modifications to the Hawking radiation's spectrum due to nonextensive statistical mechanics. In our previous paper [14], we have considered the effect of Rényi entropy and proposed that Hawking radiation originates from the effective radius $r_A$ instead of horizon $r_H$, which is the same as Giddings' proposal in finite dimension $D$. However, in that article, we still used Boltzmann-Gibbs statiscal mechanics [14]. In this article, we have applied Tsallis statistical mechanics to study the modified Hawking radiation spectrum. \\
We have obtained the modified Stefan-Boltzmann's law and modified power of Hawking radiation in $(D+1)$-dimension. We confirm the conclusion proposed by Giddings [6,7], namely, the radiation originates from the effective radius $r_A$, which extends well outside the horizon of black-hole. The lifetime of black hole and the effect of large $q$ are discussed as well. In principle, the effect of Tsallis statistical mechanics can be detected in the future and compared with the results of standard statistical mechanics.\\
Future work can be directed along at least two lines of further research. Firstly, the result in this article about Schwarzchild solution should be generalized to more generalized black-hole solutions. Secondly, general quantum field theory which is corresponding to nonextensive statistical mechanics should be applied to calculate the semi-classical Hawking radiation's spectrum.

\acknowledgments

We acknowledge beneficial discussions with Chad Briddon.


\end{document}